\documentclass[aps,prb,twocolumn,superscriptaddress,showpacs]{revtex4-1}
\usepackage{amsmath,amssymb,amsfonts,amsthm}
\usepackage{graphicx}
\usepackage{bm}
\usepackage{bbm}
\usepackage{color}
\usepackage{dcolumn}   % needed for some tables
\usepackage{epstopdf}
\usepackage[caption=false]{subfig}
\usepackage{xr}
\usepackage[colorlinks=true,linkcolor=blue,urlcolor=blue,citecolor=blue]{hyperref}
\begin{document}

 \newcommand{\breite}{1.0} %  for twocolumn

\newtheorem{prop}{Proposition}
\newtheorem{cor}{Corollary} 

\newcommand{\be}{\begin{equation}}
\newcommand{\ee}{\end{equation}}

\newcommand{\bea}{\begin{eqnarray}}
\newcommand{\eea}{\end{eqnarray}}
\newcommand{\lt}{<}
\newcommand{\gt}{>} 

\newcommand{\Reals}{\mathbb{R}}     % Reals
\newcommand{\Com}{\mathbb{C}}       % Complex #
\newcommand{\Nat}{\mathbb{N}}       % Natural #

\newcommand{\id}{\mathbboldsymbol{1}}    

\newcommand{\Real}{\mathop{\mathrm{Re}}}
\newcommand{\Imag}{\mathop{\mathrm{Im}}}

\def\O{\mbox{$\mathcal{O}$}}   % Order epsilon ... 
\def\F{\mathcal{F}}			% FourierTrafo
\def\sgn{\text{sgn}}

\newcommand{\deo}{\ensuremath{\Delta_0}}
\newcommand{\dea}{\ensuremath{\Delta}}
\newcommand{\ak}{\ensuremath{a_k}}
\newcommand{\ad}{\ensuremath{a^{\dagger}_{-k}}}
\newcommand{\sx}{\ensuremath{\sigma_x}}
\newcommand{\sz}{\ensuremath{\sigma_z}}
\newcommand{\spl}{\ensuremath{\sigma_{+}}}
\newcommand{\smi}{\ensuremath{\sigma_{-}}}
\newcommand{\alk}{\ensuremath{\alpha_{k}}}
\newcommand{\bk}{\ensuremath{\beta_{k}}}
\newcommand{\ok}{\ensuremath{\omega_{k}}}
\newcommand{\vd}{\ensuremath{V^{\dagger}_1}}
\newcommand{\vi}{\ensuremath{V_1}}
\newcommand{\vo}{\ensuremath{V_o}}
\newcommand{\zc}{\ensuremath{\frac{E_z}{E}}}
\newcommand{\xc}{\ensuremath{\frac{\Delta}{E}}}
\newcommand{\xd}{\ensuremath{X^{\dagger}}}
\newcommand{\aok}{\ensuremath{\frac{\alk}{\ok}}}
\newcommand{\tpw}{\ensuremath{e^{i \ok s }}}
\newcommand{\tpe}{\ensuremath{e^{2iE s }}}
\newcommand{\tmw}{\ensuremath{e^{-i \ok s }}}
\newcommand{\tme}{\ensuremath{e^{-2iE s }}}
\newcommand{\epls}{\ensuremath{e^{F(s)}}}
\newcommand{\emis}{\ensuremath{e^{-F(s)}}}
\newcommand{\epl}{\ensuremath{e^{F(0)}}}
\newcommand{\emi}{\ensuremath{e^{F(0)}}}

\newcommand{\lr}[1]{\left( #1 \right)}
\newcommand{\lrs}[1]{\left( #1 \right)^2}
\newcommand{\lrb}[1]{\left< #1\right>}
\newcommand{\nbt}{\ensuremath{\lr{ \lr{n_k + 1} \tmw + n_k \tpw  }}}

\newcommand{\om}{\ensuremath{\omega}}
\newcommand{\dw}{\ensuremath{\Delta_0}}
\newcommand{\wbp}{\ensuremath{\omega_0}}
\newcommand{\dv}{\ensuremath{\Delta_0}}
\newcommand{\vbp}{\ensuremath{\nu_0}}
\newcommand{\vplus}{\ensuremath{\nu_{+}}}
\newcommand{\vminus}{\ensuremath{\nu_{-}}}
\newcommand{\wplus}{\ensuremath{\omega_{+}}}
\newcommand{\wminus}{\ensuremath{\omega_{-}}}
\newcommand{\uv}[1]{\ensuremath{\mathbf{\hat{#1}}}} % for unit vector
\newcommand{\abs}[1]{\left| #1 \right|} % for absolute value
\newcommand{\norm}[1]{\left \lVert #1 \right \rVert} % for absolute value
\newcommand{\avg}[1]{\left< #1 \right>} % for average
\let\underdot=\d % rename builtin command \d{} to \underdot{}
\renewcommand{\d}[2]{\frac{d #1}{d #2}} % for derivatives
\newcommand{\dd}[2]{\frac{d^2 #1}{d #2^2}} % for double derivatives
\newcommand{\pd}[2]{\frac{\partial #1}{\partial #2}} 
% for partial derivatives
\newcommand{\pdd}[2]{\frac{\partial^2 #1}{\partial #2^2}} 
% for double partial derivatives
\newcommand{\pdc}[3]{\left( \frac{\partial #1}{\partial #2}
 \right)_{#3}} % for thermodynamic partial derivatives
\newcommand{\ket}[1]{\left| #1 \right>} % for Dirac bras
\newcommand{\bra}[1]{\left< #1 \right|} % for Dirac kets
\newcommand{\braket}[2]{\left< #1 \vphantom{#2} \right|
 \left. #2 \vphantom{#1} \right>} % for Dirac brackets
\newcommand{\matrixel}[3]{\left< #1 \vphantom{#2#3} \right|
 #2 \left| #3 \vphantom{#1#2} \right>} % for Dirac matrix elements
\newcommand{\grad}[1]{{\nabla} {#1}} % for gradient
\let\divsymb=\div % rename builtin command \div to \divsymb
\renewcommand{\div}[1]{{\nabla} \cdot \boldsymbol{#1}} % for divergence
\newcommand{\curl}[1]{{\nabla} \times \boldsymbol{#1}} % for curl
\newcommand{\laplace}[1]{\nabla^2 \boldsymbol{#1}}
\newcommand{\vs}[1]{\boldsymbol{#1}}
\let\baraccent=\= % rename builtin command \= to \baraccent
%%%%%%%%%%%%%%%%%%%%%%%%%%%%%%%%%%%%%%%%%%%%%
% End Definitions
%%%%%%%%%%%%%%%%%%%%%%%%%%%%%%%%%%%%%%%%%%%%%

%\newcommand{\IM}[1]{\texttt{\textcolor{Green} #1}}
%\newcommand{\KA}[1]{\texttt{\textcolor{RoyalBlue} #1}}
\def\red#1{{\textcolor{red}{#1}}}
%\newcommand{\IMcomment}[1]{\texttt{\color{} #1}}

%Title of paper
\title{A toy model for decoherence in the black hole information problem}

\author{Kartiek Agarwal}
\email{agarwal@physics.mcgill.ca}
\affiliation{Department of Physics, McGill University, Montr\'{e}al, Qu\'{e}bec H3A 2T8, Canada}
\author{Ning Bao}
\affiliation{Department of Physics, University of California Berkeley, Berkeley, CA , USA}
\affiliation{Computational Science Initiative, Brookhaven National Lab, Upton, NY, 11973, USA}

\date{\today}
\begin{abstract}
We investigate a plausible route to resolving the black hole information paradox by examining the effects of decoherence on Hawking radiation. In particular, we show that a finite but non-zero rate of decoherence can lead to efficient extraction of information from the evaporating black hole. This effectively pushes the paradox from becoming manifest at the Page time when the black hole has evaporated to half its size, to a timescale solely determined by the rate of decoherence. If this rate is due to a putative interaction with gravitons, the black hole at this timescale can be expected to be Planck-sized, but notably without an extensive amount of information packed inside.  We justify our findings by numerically studying a toy model of stabilizer circuits that can efficiently model black hole evaporation in the presence of decoherence. The latter is found to be well described by an effective rate equation for the entanglement and which corroborates our findings.
\end{abstract}
\maketitle

\paragraph*{\textbf{Introduction.---}} From its inception, Bekenstein \cite{bekenstein1973black} and Hawking's \cite{hawking1975particle} results that black holes emit radiation and evaporate away have posed conundrums that challenge our core physical principles and whose resolution remains a fundamental goal in modern physics. In its earliest avatar, it was realized that black hole evaporation is at odds with unitary quantum-mechanical evolution since it predicts a pure state of the black hole universe evolving into a mixed state of incoherent Hawking radiation. In its modern incarnation, it has been reformulated as an information problem by AMPS~\cite{Almheiri_2013,Almheiri_2013b}---while the effective semi-classical description valid at the black hole horizon suggests that Hawking quanta should be nearly maximally entangled with the black hole interior, an information-theoretic calculation by Page \cite{Page_1993} shows that Hawking quanta emitted by an old black hole (evaporated to half its original size), must be maximally entangled with early-time radiation. This contradicts the notion of monogamy of entanglement (or, equivalently, strong subadditivity), a fundamental result which states that a quantum subsystem cannot be maximally entangled with two different subsystems at the same time.  

%The modern incarnation of the black hole information problem \cite{Almheiri_2013,Almheiri_2013b} has been a challenging puzzle in high energy physics for much of the 2010's. At its core, it states that simultaneously requiring general relativity, quantum mechanics, effective field theory, and the "no drama" condition leads to a violation of monogamy of entanglement, as these assumptions would simultaneously require maximal entanglement between the early and late time Hawking radiation (by Page's theorem for near Haar-typical states \cite{Page_1993} and maximal entanglement between the late time Hawking radiation and the black hole interior (due to arguments about the smoothness of spacetime near the horizon, as opposed to near the singularity, of black holes.

Following AMPS~\cite{Almheiri_2013,Almheiri_2013b}, different schemes have been proposed to resolve this paradox by relaxing one of three core principles---of unitarity of quantum-mechanics, validity of effective field theory in curved geometry, and/or general relativity. These include final state projection \cite{Lloyd_2014}, the ER/EPR proposal \cite{Maldacena_2013}, state-dependent modifications of quantum mechanics \cite{Papadodimas_2013,Papadodimas_2014}, and complexity-theoretic arguments \cite{Harlow_2013,Bao_2016}.  

\begin{center}
\begin{figure*}[!htbp]
\includegraphics[width=0.9\textwidth]{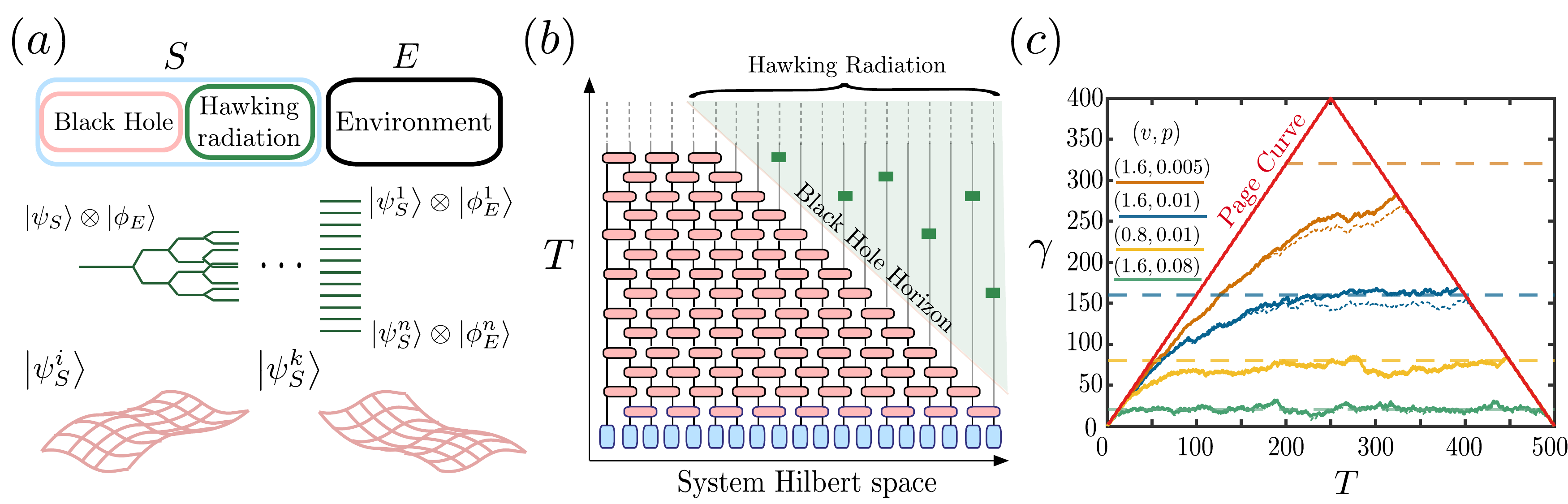}
\caption{(a) The environment, putatively of inaccessible graviton fluctuations, decoheres the system into states/'branches' with definite semi-classical geometry. Here decoherence is represented by a proliferation of basis vectors in the Schmidt decomposition of the global wavefunction, such as $\ket{\psi^i_S}$ and $\ket{\psi^k_S}$;  (b) The circuit model (see main text) captures the dynamics on a single branch of the global wave function; random two-qubit Clifford unitaries are applied inside the black hole horizon (pink) and projective measurements are made probabilistically outside (green). (c) Mutual information between the black hole and Hawking quanta (solid) in time, compared with predictions of Eq.~(\ref{eq:gammapred}) (short dashes) and steady state result of Eq.~(\ref{eq:res}) (long dashes) for different ($v$,$p$) (times are normalized to the evaporation time for $v = 1.6$). Upon contact with the Page curve, the curves subsequently follow the Page curve back down to zero at late time.}
\label{fig:mainfig}
\end{figure*}
\end{center}

Recently, techniques from decoherence in open quantum systems have been applied to this problem~\cite{Bao_2018,bao2019cosmological,campo2019decoherence}, in addition to other work connecting decoherence to black hole physics, as in \cite{Zeh_2005,Arrasmith_2019,Kiefer_2001,Hsu_2009, Demers_1996}.
 %\footnote{In particular, these decoherence based approaches are more than "mere bookkeeping;" one simply cannot shield from graviton based decoherence, for example.}. 
Within this ``operational" approach, physical ``infalling" observers able to verify the maximal entanglement between emitted Hawking quanta and the black hole cannot access certain bath degrees of freedom that decohere the global wavefunction---they effectively reside in a ``branch" of the global wavefunction which corresponds to a definite semi-classical geometry; see Fig.~\ref{fig:mainfig} (a) for illustration. If such branching due to decoherence occurs rapidly, late-time Hawking quanta grows less entangled with the early-time radiation, and more so with bath degrees of freedom; infalling observers do not witness monogamy violation because the latter is inaccessible to them~\footnote{This neatly avoids the monogamy issue, as there is no contradiction to maximal entanglement between two systems on the global wavefunction and between one of those two systems and a third on a branch of the wavefunction.}. In the context of modern approaches to the information problem, this approach is closely related both to ideas introduced in the alpha-bits work of Refs.~\cite{hayden2018learning, hayden2017approximate} and recent AdS/CFT ensemble approaches such as in Refs.~\cite{penington2019entanglement,penington2019replica, almheiri2019entropy, almheiri2019page, almheiri2019islands,almheiri2019entanglement,almheiri2019replica}. These approaches invoke a careful treatment of decoherence and do not require modifying the core principles.
In this work, we concretize the decoherence-based approach by devising a toy unitary-projective circuit model that emulates black hole evaporation and captures the decoherence of outgoing Hawking radiation; see Fig.~\ref{fig:mainfig} (b). Specifically, we consider a system of qubits which models the Hilbert space of the black hole interior and the Hawking radiation. Random Clifford gates, which form a unitary $2$-design and thus efficiently capture scrambling in the system, are applied to qubits inside of the black hole horizon which shrinks in time at the rate of $v$ qubits every time step. In the remainder of the system, projective measurements are applied with a certain probability $p$ at each time step. These measurements model decoherence of Hawking radiation by an external bath of, say, vacuum graviton fluctuations. Analysis of the mutual information $\gamma$ between early and late time Hawking quanta in this toy model, using both numerical and semi-analytical arguments, shows that the issue of monogamy violation, and thus the information paradox, becomes manifest only when the black hole has shrunk to a finite size $N^c_{\text{BH}} \approx v/p$ independent of its initial size. We use perturbative, dimensional analysis based arguments to estimate the rate of decoherence by a putative bath of gravitons, and relate the parameters $v, p$ to the physical problem of black hole evaporation. Using these results, we argue that this critical black hole size is Planckian, a scale at which quantum gravity is not fully understood, and modifications thereof could resolve the paradox. Our approach thus demonstrates the plausibility of the decoherence-based approach towards resolving the information paradox.

\paragraph*{\textbf{Circuit Model.---}} The model we consider is a unitary-projective circuit model of stabilizer states evolved in time by two-qubit Clifford gates. These Clifford gates form a unitary 2-design which efficiently describe~\cite{gottesman1998heisenberg,Hayden_2007}  information scrambling in the black hole. Although unitaries implemented by these gates are not Haar-typical with respect to all moments, they are exponentially close (in trace distance) up to the second moment, and reproduce the relevant physics with regards to Page's theorem/decoupling type arguments~\cite{Hayden_2007}.
Concretely, we study a system of $N \le 800$ qubits which begin in a random stabilizer state generated by the application of $\gtrsim N$ layers of two-qubit Clifford gates arranged in a brickwork fashion. 
For subsequent times, we assume these $N$ qubits together model both the black hole interior and its emitted Hawking radiation. The number of qubits $N_{\text{BH}}$ should be identified with the Bekenstein-Hawking entropy of the black hole, but are not meant to convey any spatial information. At each step, Clifford gates are applied in a brickwork fashion [see Fig.~\ref{fig:mainfig} (b)] to the (left) part of the chain modeling the black hole interior.  
%These Clifford gates form a unitary 2-design which efficiently describe~\cite{gottesman1998heisenberg} the conjectured information scrambling in the black hole; although unitaries implemented by these gates are not Haar-typical with respect to all moments, they are exponentially close (in trace distance) up to the second moment, and reproduce the relevant physics with regards to Page's theorem/decoupling type arguments, as argued in \cite{Hayden_2007}.
%, and which will be central to our work.
%
%Therefore, it is a robust approximation for modeling the conjectured scrambling dynamics of the black hole. 

%may itself depend on time---it is related to the scrambling time of the black hole which changes as it evaporates. For the purposes of this analysis, however, it is useful to consider fixed $T$; we will only invoke its time-dependence when specifically discussing black hole evaporation. 

We model the ejection of Hawking quanta by shrinking the black hole horizon by $v$ qubits at each time step. Since Hawking quanta emitted from physical black holes are radiated outward in all directions, we expect interactions between them can be neglected. Thus, no unitaries are applied on these qubits. Further, we assume that a local background of graviton fluctuations constantly act on and decohere these quanta. For an observer that does not have access to the graviton bath, this effect can be captured by applying projective measurements with a fixed probability $p$ on qubits representing the Hawking quanta. 

Note that steps of the random ciruit have physical meaning---$N_{\text{BH}}$ steps generate a (effectively) Haar-random state inside the black hole horizon; thus, $N_{\text{BH}}$ time steps should be associated with the scrambling time $\tau_S$ of an equivalent black hole. In particular, if we denote $T$ as the depth of the circuit, $d T / d t \approx N_{\text{BH}}/\tau_S$. In what follows, we will analyze this model ignoring any dependence of $v, p, T$ on physical time $t$ and re-invoke this time dependence when directly relating results to black hole evaporation. Also, although one may track the density matrix corresponding to the full ensemble of measurement outcomes, it is sufficient to track a specific (randomly chosen) outcome for the purpose of evaluating the \emph{mutual information} between the black hole interior and the radiated Hawking quanta~\footnote{An observer on a particular branch sees a particular outcome of measurement. For such an observer, entanglement between Hawking radiation and the black hole interior is identical to the mutual information. Also, note that the mutual information change does not depend on the specific outcome of the measurement.} Henceforth we equivalently refer to the above as mutual information \emph{across the horizon}. 

%We now remark on a few subtleties. 
%First, the two-qubit Clifford gates we employ form a unitary 2-design and therefore form a finite set of two-qubit gates that correctly model information scrambling in an equivalent circuit model in which the gates are drawn from the Haar measure. The advantage of working with these gates is that the evolution of the wavefunction under the action of such operations can be modeled efficiently with $\sim \mathcal{O} \left( N^3 \right)$ computations~\cite{gottesman1998heisenberg}. This allows us to work with large enough systems to glean the physics which we discuss below. Second, 
%First, with regard to measurements, 

%When the density matrix of all possible measurement outcomes is considered, the entanglement entropy we compute is more generally given by the mutual information across the horizon. With this fair warning, we will use the former term henceforth. 

\begin{center}
\begin{figure}[!htbp]
\includegraphics[width=3in]{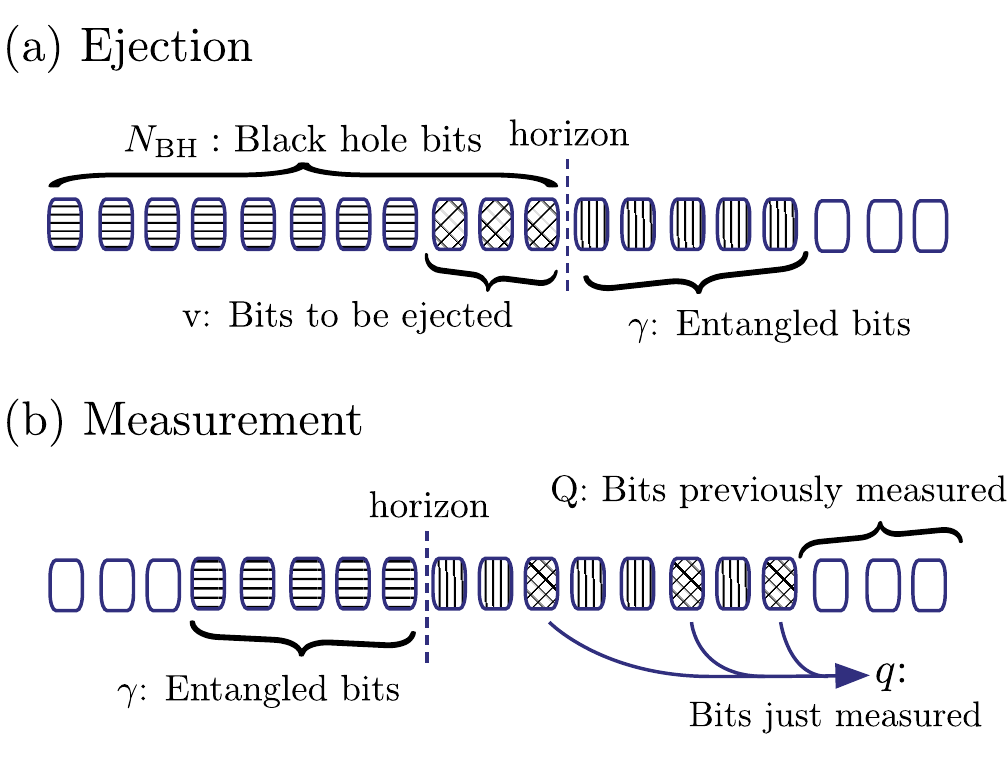}
\caption{Processes of black hole evaporation, with bits organized to facilitate the application of the decoupling theorem.}
\label{fig:bhprocesses}
\end{figure}
\end{center}

Before proceeding further, we reiterate the statement of the information paradox in this setting. Our model captures black hole evaporation in the Hilbert space accessible to a physical infalling observer. %To avoid the information paradox, 
This observer must find near maximal entanglement between a newly emitted Hawking quanta and the black hole interior to be consistent with effective field theory calculations at the horizon. The decoupling theorem~\cite{Page_1993,dupuis2010decoupling} states---see Fig.~\ref{fig:bhprocesses} (a) for illustration--that this will be the case as long as the number of qubits inside the black hole exceeds the mutual information across the horizon by a finite amount. (The entanglement is then exponentially close in this finite amount to being maximal.) Thus, within our scheme, the information paradox does not occur until the mutual information nearly equals $N_{\text{BH}}$.  Equivalently, the paradox becomes manifest when the mutual information approaches the Page curve (given by the maximal possible entanglement between the black hole and radiated Hawking quanta, $\text{min.} \left[N_{\text{BH}}, N-N_{\text{BH}} \right]$). Measurements of the Hawking quanta reduce this mutual information and can dramatically delay the onset of the paradox.

\paragraph*{\textbf{Motivation of Assumptions.---}}
Here we provide a motivation for the key assumptions made in our toy model, namely the split between the Hilbert space of the Hawking quanta and the decohering environment, and the lack of access to this environment by physical observers. 

Let us, for simplicity, consider a black hole in one spatial dimension. As it radiates Hawking quanta to the left, it is kicked a little to the right, and vice versa. The entire wavefunction of the black hole and the surrounding universe thus exists in a superposition of different spacetime geometries described by the spatial distribution of Hawking quanta. At early times, these geometries do not differ significantly. However, the information paradox usually sets in when about half the black hole has been radiated. Thus, the spacetime geometries in the superposition can become macroscopically different at the Page time. 

Now, in much the same way a physical observer observes one particular orientation of the magnetic moments in a macroscopic magnet, a physical observer observing such a universe will lose access to spacetime geometries that are inconsistent with their (local) measurements. In this sense, the observers of the black hole and Hawking radiation will observe a projected wave-function which comprises of a superposition of states and geometries consistent with the measurements of this observer.   

It is also important to consider how the above unfolds in regards to the evolution of the entanglement structure of this system. It is the above situation which we attempt to model. Let us assume the above black hole exists in a purely gravitational universe. The wave-function of the Hawking radiation and the black hole then evolves and interacts with gravitons. These gravitons can, for instance, be sourced from the measurements being made by the observer, but there is also intrinsic decoherence from thermal (or even vacuum) graviton fluctuations. Assuming the Hawking quanta is completely entangled with the black hole interior initially, this interaction (described by a unitary process) \emph{transfers} the entanglement to that between the black hole and the gravitons. Now, if an observer could uncover this macroscopic amount of entanglement between the black hole and the gravitons, they would, in principle, be conscious of the fact that they exist in a superposition of multiple spacetime geometries. Thus, via the same arguments as given above, it is reasonable to assume that a physical observer cannot, in fact, detect this entanglement beetween the black hole and graviton fluctuations. 

Our circuit model effectively describes presicely this process of Hawking quanta getting entangled with a decohering environment of inaccessible gravitons. This undetectable environment can be therefore modeled simply as measurements of the Hawking quanta. This completes our justification of our circuit model and key assumptions. 

\paragraph*{\textbf{Numerical Results.---}} Some results for different choices of (fixed-in-time) $v, p$ are presented in Fig.~\ref{fig:mainfig} (c).  The results exhibit a simple trend: mutual information across the horizon grows to approximately the value $\sim v/p$ provided the system is large enough, that is, $N_{\text{BH}} \gg v/p$; eventually, the mutual information tracks the Page curve. We now derive an approximate equation of motion for entanglement growth that captures the steady state result $\gamma \approx v/p$ and well approximates the dynamical results. 

\paragraph*{\textbf{Equation of motion for the mutual information.---}}
To describe entanglement growth, it is first important to visit the decoupling theorem and its adaptation to the present setting; see Ref.~\cite{choi2019quantum} for a similar discussion. Consider two coupled systems $A$ and $B$ initialized in a joint pure state. Assume A(B) has $\gamma$ bits of (mutual) information about B(A). We can perform a unitary transformation that acts on $B$ and distills the $\gamma$ maximally entangled bits in $B$ into a subsystem $\tilde{B}$. By monogamy, the rest of $B$ is now unentangled with $A$ and $\tilde{B}$. The central question we wish to answer is---how much information does measuring $N_{\text{m}}$ qubits in $A$ yield about $\tilde{B}$ (and thus B)? Assuming system $A + \tilde{B}$ is in some Haar-typical state, the decoupling theorem states, that if qubits $N_{\text{m}} + \gamma <  N_A - N_{\text{m}}$, then the qubits $N_{\text{m}}$ have exponentially little information about the $\gamma$ entangled qubits in $\tilde{B}$~\footnote{This startling result is natural when one considers monogamy of entanglement---in a Haar-typical state, the $N_{\text{m}} + \gamma$ bits are maximally entangled with the larger subsystem of $N_A - N_{\text{m}}$ bits which makes it impossible for the $N_{\text{m}}$ measured bits to have information about the $\gamma$ bits in $B$.}. 

Readers familiar with the Eigenstate Thermalization Hypothesis (ETH) can intuit this result by noting that the reduced density matrix corresponding to a spatially local region in a pure state, in a system with no conservation laws, is well approximated by the identity matrix, until this local region grows close to half the system size. When the reduced density matrix is the identity matrix, bits within this local region are clearly unentangled with each other, and are rather maximally entangled with the outside. Here we consider the `local region' to comprise of the bits measured and the distilled bits of $\tilde{B}$. If this system comprises less than half the size of the total system, the measured bits and the distilled bits together are represented by the identity matrix, and are thus entirely unentangled with one another. 

This result will be used repeatedly below to formulate dynamics of the mutual information. 
%In the opposite situation, when $N_{\text{m}} + \gamma > N_A - N_{\text{m}}$, the measured bits do have information about subsystem $B$; in this case, we can make a mean-field like assumption to quantify the information per bit in $A$ about subsystem $B$---it is simply given by $\gamma / N_A$. The above intuition will be central to the development of the entanglement equation of motion. 

We now discuss the rules for entanglement change $\Delta \gamma (T)$ as a function of the current entanglement $\gamma (T)$, the bits to be ejected in this time step $v(T)$, the present number of black hole qubits $N_{\text{BH}}(T)$, unique qubits $q(T)$ measured in this time step, and $Q(T)$, which counts all unique qubits measured at times before the present time $T$. See Fig.~\ref{fig:bhprocesses} for an illustration of these quantities. 

Two processes occur that change the mutual information: ejection of qubits from the black hole, and measurement of qubits outside the black hole. Note that measurement eliminates the measured qubit from further participating in the dynamics. In what follows, we will make the assumption that all these processes keep the wavefunction of the system in a Haar-typical state of appropriate dimensionality (accounting for the entanglement across the horizon, and the qubits removed by measurement)---thus the decoupling theorem will always apply on an appropriately defined set of qubits in the system. In Fig.~\ref{fig:bhprocesses}, this corresponds to the qubits drawn with some form of patterning. 

During ejection, if $N_{\text{BH}} - v > v + \gamma$, the decoupling theorem implies that $v$ qubits to be ejected are monogamously entangled with the rest of the black hole interior. Thus, each such ejected qubit contributes an increase $\Delta \gamma \approx +1$. If the former condition is not satisfied, then the ejected qubits contain little to no information about the qubits inside the black hole. In the extreme case when they are maximally entangled with the black hole exterior, their ejection results in a reduction of entanglement $\Delta \gamma = -1$ per qubit. The latter occurs when $\gamma(T)$ approaches its maximal value $N_{\text{BH}}(T)$. Assuming $v(T)$ is a small number, one may capture the above approximately by

\be
\Delta \gamma_{\text{ej}} = v\cdot \text{sgn}  \left( N_{\text{BH}} - \gamma - 2 v \right)
\label{eq:dgamej}
\ee 

where $\text{sgn} (x)$ is the sign function. 

During measurement, if $q + \gamma < N - N_{\text{BH}} - Q - q$, then the measured qubits reveal (exponentially) close to nothing about the black hole's internal structure. As a result, mutual information across the horizon is changed only if the former condition is violated. In that event, we make a mean field-like assumption that the information $\gamma$ is distributed equally among the $N - N_{\text{BH}} - Q$ external qubits that have not already been measured. Subsequently, each measured qubit decreases entanglement across the horizon by an amount $\gamma / \left( N  - N_{\text{BH}} - Q\right)$. Thus,  

\be
\Delta \gamma_{\text{me}} = - q \cdot \frac{\gamma }{ N - N_{\text{BH}} - Q} \cdot \Theta \left( 2 q + \gamma - N + N_{\text{BH}} + Q  \right)
\label{eq:dgamme}
\ee

where $\Theta$ is the Heaviside function. The change of entanglement then is given by the sum of the two contributions---
\be
 \Delta \gamma (T) = \gamma (T+1) - \gamma(T) = \Delta \gamma_{\text{ej}} (T) + \Delta \gamma_{\text{me}} (T)
 \label{eq:gammapred}
\ee

\begin{table}[!htbp]
\centering
 \begin{tabular}{||c | c | c||} 
 \hline
  & Black Hole & Random Circuit \\ [0.5ex] 
 \hline\hline
  Time & $t$ & \text{Circuit depth} $T$  \\
 \hline 
 Black Hole DOF & $ S_{\text{BH}} = 4 \pi M^2 $ & Bits inside horizon, $N_{\text{BH}}$  \\ 
 \hline
 Scrambling Time & $\tau_S = \kappa M$ &  Circuit of depth $N_{\text{BH}}$  \\
 \hline
% Evaporation Rate & $\d{M}{t} = - \frac{1}{15360 \pi M^2}$ & $\Rightarrow \d{N_{\text{BH}}}{t} \sim \frac{-\sqrt{\pi}}{960 \sqrt{N_{\text{BH}}}} , \; \d{T}{t} \sim \frac{\kappa}{3840 N^2_{\text{BH}}} \; \Rightarrow \d{N_{\text{BH}}}{T} \equiv v \approx - \frac{4\sqrt{\pi}}{\kappa} \left(N_{\text{BH}} \right)^{3/2} $  \\
Evaporation Rate & $\d{M}{t} = - \frac{b}{M^2}$ & $v \equiv \d{N_{\text{BH}}}{T} \sim N_{\text{BH}}^{-1} $  \\
 \hline
 Measurement Rate & $\d{\bar{p}}{t} = a \left(\frac{1}{8 \pi M}\right)^x$ & $p \equiv \d{\bar{p}}{T} \sim  N_{\text{BH}}^{-(x+1)/2} $  \\  
 \hline
\end{tabular}
 \caption{ A mapping between the black hole evaporation problem and the stabilizer circuit model of Fig.~\ref{fig:mainfig} (b). The equivalence of the scrambling time of the black hole, and that of the random circuit is the constitutive relation tying the flow of time in the physical problem to circuit depth; it asserts $\d{T}{t} = \frac{N_{\text{BH}}}{\tau_S}$. The remaining relations are found by simple algebraic manipulations and well-established black hole physics.} %The objects on the left-hand side are well-known results from black hole physics, while the objects on the right are their analog in the stabilizer circuit. }
 %The first row establishes the relationship between the number of qubits inside the black hole and the number of qubits still acted on by the random Clifford unitaries. The second row relates the black hole scrambling time to its effective analog in the spin chain, as governed by the brickwork time $T$. Here we upper bound the logarithm by a sufficiently large constant. The third row relates the evaporation rate of the black hole to the modeled evaporation rate $v(t)=\d{N_{\text{BH}}}{T}$. In obtaining this, we used that $\d{N_{\text{BH}}}{t}$ scales like$ \frac{-\sqrt{\pi}}{960 \sqrt{N_{\text{BH}}}}$ from black hole physics; further, the transformation from the brickwork time to real time is given by $\d{T}{t}=\frac{\kappa}{3840 N^2_{\text{BH}}} $.Putting these together yields the stated result of $\d{N_{\text{BH}}}{T} \equiv v \approx - \frac{4\sqrt{\pi}}{\kappa} \left(N_{\text{BH}} \right)^{3/2}$. Finally, in the last row, we relate the measurement rate in real time to the measurement rate in the brickwork time. The measurement rate in real time is worked out using dimensional analysis argument sin the main text, while the conversion to brickwork time follows from another simple application of $\d{T}{t}$.}
  \label{tab:relations2}
\end{table}
The above equations are easily solved in the situation where the entanglement entropy remains small compared to the size of the black hole, $N_{\text{BH}}$. In this case, $\Delta \gamma_{\text{ej}} \approx v$---thus, each ejected qubit is well entangled with the black hole interior (as is necessary for consistency with effective field theory). 
Eq.~(\ref{eq:dgamme}) can be simplified, but requires a self-consistent assumption. Suppose that throughout the evolution of the system, the Hawking quanta are maximally entangled with the black hole interior. (This will not hold true, for instance, if the rate of measurement was zero.) When measured, these Hawking quanta lose entanglement with the black hole interior and the mutual information across the horizon decreases by the number of Hawking bits measured. In the above picture, $\gamma \approx N - N_{\text{BH} - Q}$, and one notes that the argument of the Heaviside theta function  in Eq.~(\ref{eq:dgamme}) is $2q > 0$. We also note $\avg{q/(N-N_{\text{BH}}-Q)}=p$ (the average number of bits measured, $q$ is given by the Hawking quanta remaining, $N- N_{\text{BH}} - Q$ multiplied by the probability of measurement, $p$). Thus, we find  $\Delta \gamma_{\text{me}} \approx \gamma \cdot q / \left( N - N_{\text{BH}} - Q \right) = \gamma \cdot p$. 

Note that the above simplification was obtained starting with the assumption that the emitted Hawking quanta are maximally entangled with the black hole interior, which is equivalent to assuming $N_{\text{BH}} > \gamma + 2 v$ by the decoupling theorem [see also Eq.~(\ref{eq:dgamej})] which is in turn true provided the entanglement remains area law ($\gamma \ll N_{\text{BH}}$) and the rate of ejection of bits, $v$, is finite. The latter is true by design, and we will see that the assumption of area law entanglement is justified self-consistently. Putting the above together, we find that the evolution of $\gamma$ is described by a straightforward detailed balance equation

\be
\d{\gamma}{T} = v - \gamma \cdot p \;; \; \gamma_{\text{area law}} = \frac{v}{p}
\label{eq:res}
\ee

where i) we immediately see that the steady state solution satisfies the area-law assumption for $\gamma$, and ii) it matches nicely with our numerical results, see Fig.~\ref{fig:mainfig} (c). Note that the solution is naturally stable to perturbations corresponding to blips of enhanced/lowered rate of measurement of Hawking quanta---these perturbations are damped over a circuit depth $\sim 1/p$. When $v, p$ are time dependent, as in the case of black hole evaporation, for the solution to remain instantaneously valid, we require the rate of change $\abs{\d{\text{log} \left( v/p \right)}{T}} \ll p $ or $\abs{ \d{(v/p)}{T} } \ll v$. 

%In the simulations below, we will in fact set $v(T)$ and $p(T)$ to be independent of time $T$, instead focussing on the critical properties of the 
\paragraph*{\textbf{Implications for black hole evaporation.---}} We first discuss the mapping of the circuit model to black hole evaporation as enumerated in Table~\ref{tab:relations2}; below we use natural units $\hbar = c = G = 1$. The circuit depth $T$ is the natural analog of physical time $t$. The Bekenstein-Hawking entropy $S_{\text{BH}} = 4 \pi M^2$ enumerates black hole degrees of freedom (DOF); thus it must be given by the $N_{\text{BH}}$ qubits inside the horizon in the circuit model. The black hole scrambling time $\tau_S \sim \mathcal{O} \left( M \log M \right) \le \kappa M$ (where we have subsumed the weak log-dependence on $M$ into a sufficiently large constant $\kappa$) corresponds to a circuit of depth $N_{\text{BH}}$ which completely scrambles (transform into Haar-random state) any initial state of the $N_{\text{BH}}$ black hole qubits. Note that the equivalence of the scrambling time in the two models gives physical meaning to the circuit depth; in particular, it asserts in the continuum limit the relation
\be
\d{T}{t} \approx \frac{N_{\text{BH}}}{\tau_S} = \sqrt{\frac{N_{\text{BH}}}{4 \pi \kappa^2}}
\label{eq:evaprel}
\ee 

The Hawking result for the evaporation rate can be translated to the number of qubits ejected $v \equiv d N_{\text{BH}} / d T \sim N_{\text{BH}}^{-1}$ in a brickwork time step using preceding relations. To determine the measurement rate, we consider the following minimal setting for decoherence---imagine a Schwarzchild black hole in four dimensional flat space at zero temperature wherein Hawking quanta is decohered only by vacuum graviton fluctuations. There are two relevant energy scales for such a process---the Hawking temperature $T_H = 1/(8 \pi M)$, and the Planck temperature. The latter is $1$ in the Planck units we work in. Note that in a more realistic setting, there will be additional sources of noise, which will correspond to an increased rate of decoherence as compared to our estimate in this minimal setting; thus, we do not expect our main conclusions to change. 

Now, in the minimal setting we consider, the decoherence rate $d \bar{p} / dt $ is some undetermined function of $T_H$ in natural units. Assume then $\d{\bar{p}}{t} = a T^x_H = a \left( \frac{1}{8 \pi M} \right)^x$ with an exponent $x \sim \mathcal{O} (1)$, and undetermined $\mathcal{O} \left( 1 \right)$ constant $a$. The probability of measurement per time step in the random circuit is then given by $p \equiv d\bar{p}/dT \sim \left(N_{\text{BH}}\right)^{-(x+1)/2}$. Finally, a naive Fermi's Golden Rule calculation suggests $d \bar{p} / d t \sim T^2_H$ or $x = 2$---the dependence comes from the square of the coupling proportional to the temperature $T_H$ of Hawking radiation to gravitons, which in turn have an intrinsic density of states independent of $T_H$; see also the result of Ref.~\cite{bao2019cosmological}. Note that the above functional form may be considered to apply to the most relevant term in a Taylor expansion in $T_H$ of the decoherence rate around $T_H = 0$. Terms higher order in $T_H$ are not precluded from the analysis, but they do not impact our main conclusions.  

Given Table~\ref{tab:relations2}, and for $x < 3$ (valid for the Fermi's Golden Rule estimate), we find the ratio $v/p \ll N_{\text{BH}}$, and the stability condition $\abs{ \d{(v/p)}{T} } \ll v$ is satisfied for a large enough black holes. In fact, the two conditions amount to the same relation barring an $\mathcal{O} \left( 1 \right)$ constant. Thus, we anticipate for evaporating black holes, Eq.~(\ref{eq:res}) applies and a local continuous-in-time decoherence of Hawking radiation is enough to keep entanglement instantaneously tracking the value $v/p$ until it shrinks to a critical size $N^c_{\text{BH}}$ at which point the mutual information across the horizon saturates its internal degrees of freedom. 

%(We note that the requirement $x < 3$ only applies to the functional dependence of the most relevant term in the limit $T_H \rightarrow 0$; terms at higher orders in a Taylor expansion of the decoherence rate will become more relevant as the black hole continues to evaporate and $N_{\text{BH}} \rightarrow 0$.)

Until this critical limit is reached, the entanglement remains below the Page curve, and as per Eq.~(\ref{eq:dgamej}), emitted Hawking quanta are maximally entangled with the black hole, thus avoiding monogamy violation. Beyond this limit, the black hole is evaporating faster than information can be extracted from it, and newly emitted Hawking quanta appear to be maximally entangled with the exterior, signalling the onset of the information paradox. However, as mentioned above, this critical black hole has a size $N^c_{\text{BH}}$ that is Planckian; in particular

\be
N^c_{\text{BH}} = \left( \frac{ 16 \pi^{3/2} \left( 4 \sqrt{\pi}\right)^x b }{a} \right)^{\frac{2}{(3-x)}}. 
\label{eq:criticalblackhole}
\ee

in terms of undetermined $\mathcal{O} \left( 1 \right)$ constants $a, b,$ and $x$; note that the scrambling time in fact drops out of the consideration completely, marked by an absence of $\kappa$ .

\paragraph*{\textbf{Discussion.---}} As mentioned above, Eq.~(\ref{eq:criticalblackhole}) suggests that the black hole information paradox becomes an issue only when the black hole is Planck-sized. At this point, the paradox may be resolved invoking currently unexplored Planck-scale physics, where a myriad of possibilities may occur---for instance, the black hole may simply stop evaporating~\cite{Chen_2015} (importantly, this remnant black hole need only store information within the limit of the Bekenstein bound). It is however, important to remind ourselves that this result has been obtained strictly under the assumption that there exists a sensible distinction between the Hilbert space corresponding to Hawking quanta and the decohering environment, and that physical observers do not have access to this environment. The validity of these assumptions is not by any means obvious. However, our work provides a strong impetus to consider further exploration into such ideas.  

We note that the decoherence rate is proportional to the ratio of the two dimensionful scales in the problem, namely the black hole mass/temperature scale and the Planck scale. As there are only two dimensionful scales in the problem, only their ratio that provides a dimensionless scalar can enter into the scalar probability rate. We have thus far worked in natural units where the Planck scale has been set to 1, thus the lack of explicit appearance of the Planck scale. In particular, the power that is to be taken is driven by an application of Fermi's Golden Rule, which suggests the power of two in the rate scaling.

While the applicability of this work to black hole evaporation is highly speculative, we anticipate these results could be used to motivate experiments on noisy intermediate-scale quantum computers~\cite{preskill2018quantum} and other artificial quantum systems~\cite{islam2015measuring,kaufman2016quantum,linke2018measuring} to specifically study quantum entanglement dynamics and benchmark their progress. The physics we explore here is most clearly seen in spin chains much larger in size than those than can be simulated via exact diagonalization, but which do not require scaling to the thermodynamic limit.  

\paragraph*{\textbf{Note Added.---}} During the completion of this work, we learned of upcoming work \cite{Qiunpublished} that could be synergistic with the ideas in this work.

\paragraph*{\textbf{Acknowledgements.---}} KA and NB acknowledge fruitful discussions with Ehud Altman, Simon Caron-Huot, Aidan Chatwin-Davies, Newton Cheng, Soonwon Choi, Michael Gullans, Prahar Mitra, Jason Pollack, Brian Skinner, Vincent Su. We would especially like to thank Aidan Chatwin-Davies and Jason Pollack for providing comments on the draft. NB is particularly grateful for a seminar given as part of a GeoFlow collaboration workshop by Soonwon Choi that introduced him to unitary-projective random circuits. We would also like to acknowledge the comments of an anonymous referee, for strengthening the motivations and detailed arguments presented in this work. KA likewise is particularly grateful to Michael Gullans for introducing him to and providing many clarifications on various aspects of random stabilizer circuits and related quantum information theory. KA acknowledges support from NSERC Grants RGPIN-2019-06465, and DGECR-2019-00011. NB is supported by the National Science Foundation under grant number 82248-13067-44-PHPXH, by the Department of Energy under grant number DE-SC0019380, and by New York State Urban Development Corporation Empire State Development contract no. AA289. 

\bibliographystyle{apsrev4-1}
\bibliography{stuff.bib}

\end{document}